\RequirePackage{fix-cm}

\documentclass[pdftex,epjc3]{svjour3}
\RequirePackage[T1]{fontenc}

\smartqed

\RequirePackage{graphicx}
\RequirePackage{mathptmx}
\RequirePackage[numbers,sort&compress]{natbib}
\RequirePackage[colorlinks,citecolor=blue,urlcolor=blue,linkcolor=blue]{hyperref}

\journalname{Eur. Phys. J. C}

\begin{document}

\title{Signature of biased range in the non-dynamical Chern-Simons modified
gravity and its measurements with satellite-satellite tracking
missions:Theoretical studies}

\author{Li-E Qiang\thanksref{e1,addr1} \and Peng Xu\thanksref{e2,addr2,coor}}

\thankstext{e1}{qqllee815@chd.edu.cn}
\thankstext{e2}{xupeng@amss.ac.cn}

\thankstext{coor}{Corresponding author.}

\institute{Department of Geophysics, College of
Geology Engineering and Geomatics, Chang'an University, Xi'an, 710054,
China,\label{addr1}\and Academy of Mathematics and Systems Science,
Chinese Academy of Sciences, Beijing, 100190, China. \label{addr2}}


\maketitle

\begin{abstract}
Having great accuracy in the range and range rate measurements,
the GRACE mission and the planed GRACE Follow On mission can in principle
be employed to place strong constraints on certain relativistic gravitational
theories. In this paper, we work out the range observable of the non-dynamical
Chern-Simons modified gravity for the Satellite-to-Satellite
Tracking (SST) measurements. We find out that a characteristic time accumulating
range signal appears in non-dynamical Chern-Simons
gravity, which has no analogue found in the standard parity-preserving
metric
theories of gravity. The magnitude of this Chern-Simons range signal will reach
a few times of $\chi cm$ for each free flight of these SST
missions, here $\chi$ is the dimensionless post-Newtonian parameter of the
non-dynamical Chern-Simons theory. Therefore, with the 12 years data of the
GRACE mission, one expects that the mass scale
$M_{CS}=\frac{4\hbar c}{\chi a}$ of the non-dynamical Chern-Simons gravity could
be constrained to be larger than $1.9\times10^{-9}eV$. For the GRACE FO mission
that scheduled to be launched in 2017, the much
stronger bound that $M_{CS}\geq5\times10^{-7}eV$ is expected.
\keywords{Chern-Simons Theories, Classical Theories of Gravity, Models of
Quantum Gravity}
\end{abstract}

\section{Introduction and motivations \label{sec:intro}}

Einstein's general theory of relativity (GR), as the fundamental
theory for gravitation and dynamical spacetime, is
one of the corner stones of modern physics and cosmology.
From the late 1960s, with the establishment of the Dicke framework
\cite{Dicke1964} and the parameterized post-Newtonian (PN)
formalism \cite{Nordtvedt1968,Will1971,Will1972,Will1973},
GR had passed many stringent tests with scales ranging
from $1mm$ to $1kAU$ \cite{Turyshev2008,Will2014}.
While, recently, observations from astrophysics and cosmology had
given rise to new challenges to GR,
which are known as the dark matter and dark energy problems
\cite{Navarro1996,Peebles2003,Bertone2005,Clowe2006,Li2013}. Concerning
these, different classes of modified
gravitational theories had been developed,
please consult \cite{Clifton2012} for details. On the other hand, in searching
for the union of quantum mechanics and gravity, modifications and extensions to
GR also arose naturally along the different approaches to this ``Holy-Grail'' of
fundamental physics.

Among the modified theories, the extensions to the
Einstein-Hilbert action with second order curvature terms are of particular interest,
which may arise in the full, but still lacking, quantum theory of
gravity as high energy corrections to GR, see \cite{Niedermaier2006}.
The string theory inspired Chern-Simons (CS) modified gravity
\cite{Deser1982,Campbell1990,Campbell1991,Jackiw2003,Alexander2009},
with the additions of a parity-violating term $R\;^{\star}R$ and
a coupling scalar field $\theta$, is one of such extensions.
CS modified gravity appeared first in \cite{Deser1982} as a (2+1)-dimensional
model, and then it was extended into 3+1
dimensions as a consequence of the string theory
\cite{Campbell1990,Campbell1991}. Now, it
is known that CS gravity is required by all 4-dimensional compactifications
of string theory for self-consistency \cite{Polchinski1998}. Being a
promising model, CS modified gravity has found connections
with different fields such as gravitational physics, particle physics,
string theory, loop quantum gravity, and cosmology, please consult
\cite{Alexander2009} for detailed discussions.

CS modified gravity now contains two classes of formulations,
the non-dynamical and dynamical formulations, which are in fact two
distinct theories. In the non-dynamical formulation, the CS scalar $\theta$
is externally prescribed, which depends on the mass scale $M_{CS}$ of the
specific theory under the consideration.
while, in the more realistic but complicate dynamical formulation,
the evolution of the CS scalar is then sourced by the spacetime curvature.
The non-dynamical CS gravity now serves as a useful model that provides us
insights into parity violating theories of gravity. Up to now, the
tests and constraints on both the non-dynamical and dynamical CS gravity
are all based on the observations from astrophysics and space
based experiments. The first but
weak bound on the CS scalar
$\theta$ or the mass scale was obtained in \cite{Smith2008}
 based on the results from the
LAGEOS I, II \cite{Ciufolini2004,Ciufolini2007,Ciufolini2010a} and the Gravity
Probe-B \cite{Everitt2011} missions, which had placed the constraint
$M_{CS}\geq2\times10^{-13}eV$. In the attempts to explain the
flatness of galaxy rotation curves, new bound was also obtained
\cite{Konno2008}. The strongest bound on the CS mass scale up to now was
based on the data from double binary pulsars \cite{Yunes2009a,Ali-Haimoud2011a},
which had the constraint $M_{CS}\geq 4.7\times10^{-10}eV$ as been revised in
\cite{Ali-Haimoud2011a}. For the tests of the dynamical CS gravity,
the studies became active only recently. The vacuum
solutions outside rotating black holes and stars in the dynamical formulation
were studied with the slow rotation approximation
 \cite{Yunes2009b,Ali-Haimoud2011,Yagi2012a}, and their possible
tests can be found in \cite{Chen2010,Yagi2013,Vincent2013}. Moreover, the
parity-violating term $R\;^{\star}R$ also leaves
distinguishable signatures in gravitational waves, which may be captured by
ground based or future space borne gravitational wave
antennas \cite{Sopuerta2009,Garfinkle2010,Pani2011,Canizares2012}.

Based on the PN analysis of the non-dynamical CS gravity in
\cite{Alexander2007,Alexander2007a},
we suggest here a new method to place a rather strong constraint on the
mass scale of the non-dynamical theory. A characteristic range
observable $\delta \rho^{CS}$ is found here for the non-dynamical theory which
could be measured by the operating and
future planned Satellite-to-Satellite Tracking (SST) missions,
that the GRACE and GRACE Follow On (GRACE FO) missions.
To summarize here
\begin{equation}
\delta\rho^{CS}=-\frac{\chi G J\rho_{0}\sin
  i}{2c^2a^3}(\sin(\omega t) t+\frac{\cos(\omega
t)}{\omega}),\label{eq:signal}
\end{equation}
where $\chi$ is the new PN parameter of the non-dynamical theory
\cite{Alexander2007}, and
$J$ denotes the Earth angular momentum. $i$, $\omega$,
 and $a$ denote the inclination,
angular frequency and semi-major of the orbit of the GRACE
or GRACE FO satellites, and $\rho_{0}$ denotes
the averaged range between the two satellites. The key result
turns out to be that the CS range observable contains an oscillating term that
growing
linearly with time, which has no
analogue found in standard parity-preserving metric theories.
In each free flight of these SST missions, $\delta\rho^{CS}$
will reach to a few $\chi cm$.  With the 12
years data from the GRACE mission, one expects that the mass scale of the
non-dynamical CS gravity will be constrained to $M_{CS}\geq1.9\times10^{-9}eV$.
For the future GRACE FO mission, an even stronger bound, that
$M_{CS}\geq5\times10^{-7}eV$, is expected. Therefore,
in principle, with the help of these SST missions, one could place, up to now,
the strongest constraints on the CS modified gravity.

This paper focuses on the theoretical studies and expands as follows. We first
give a brief introductions on the status of the
GRACE and GRACE FO missions in section.\ref{sec:GRACE}. The non-dynamical CS
modified gravity is briefly reviewed in section.\ref{sec:NCS}. The detailed
derivations of our results is described in section \ref{sec:CS}. Finally, we
discuss the measurements of the CS range signal with the GRACE and
GRACE FO missions in section.\ref{sec:measurements}. As mentioned
before the non-dynamical CS gravity can only serves as a model mimic
the the more realistic dynamical one. The studies of the range observable in
dynamical CS
gravity in the slow rotation approximation will
be left in future works.

\section{GRACE and GRACE Follow On missions \label{sec:GRACE}}
The Gravity Recovery And Climate Experiment (GRACE) mission is a joint
mission between the National Aeronautics and Space Administration
(NASA) in the United States and the Deutsche Forschungsanstalt f\"{u}r Luft und
Raumfahrt in Germany, which was launched in March of 2002
\cite{Tapley2004}.
The aim of GRACE mission is to accurately map the variations of Earth gravity
field for a nominal mission lifetime of five years.
Today, GRACE is still operating in an extended mission phase, which is expected
to continue through at least 2015
\footnote{Please see \url{http://www.csr.utexas.edu/grace/},\\
\url{http://www.nasa.gov/mission_pages/Grace/index.html}
and\\
\url{
https://earth.esa.int/web/guest/missions/3rd-party-missions/current-missions/grace}.}.
GRACE is a SST mission at low Earth orbit, which
is consisted of two identical
satellites that following almost the same near circular polar
orbit one after another,
please see table.\ref{tab:GRACE} for the samples of the orbit elements.
\begin{table*}
 \begin{center}
 \begin{small}
\begin{tabular}{lllllll}\hline
S/C & $a$ [km] & $e$ & $i$ [degree] & $\Omega$ [degree] & $\omega$ [degree] &
$\mathcal{M}$ [degree]\\
\hline
GRACE A & 6841.11877& 0.00272831 & 89.9395 & -71,5742 & 119.916 & -179,997\\
\hline
GRACE B & 6839.80210 & 0.00298412 & 89.8374 & -71.5081 & 118.082 & -179.997\\
\hline
 \end{tabular}
 \end{small}
 \end{center}
\caption{Samples of the GRACE orbit elements, which are the
semi-major axis $a$, the eccentricity $e$,
the
orbital inclination $i$ to the Earth equator, the longitude of the ascending
node $\Omega$,
the argument of pericenter $\omega$, and the mean anomaly $\mathcal{M}$. The
Keplerian
orbital periods of the GRACE pair are of the order of $1.56 h$. This data is
from 13 September 2003.
}
\label{tab:GRACE}
\end{table*}
The two satellites are separated along-track from each other by
$170km\sim270km$ maintained by occasional orbit
maintenance manoeuvres, and linked continuously by highly
accurate inter-satellite K-Band Ranging system. The SST measurement has
the accuracy about
$10\mu m/\sqrt{Hz}$ for biased range and about $1\mu m/s\sqrt{Hz}$ for range
rate in the signal band of $10^{-2}Hz\sim 10^{-1}Hz$ \cite{Kim2002,Tapley2004}.
Near the orbital frequency $\sim 10^{-4}Hz$, the accuracy in the range
measurement is still about $1cm\sim 2cm$ \cite{Kim2002}.
To be brief, such high accuracy is obtained by the combination of the
dual-frequency one-way K-Band
phase measurements carried on each satellite, which can largely remove the
noises from the instability of the on-board ultra stable oscillators and
errors from Earth ionosphere. Also, GRACE carries accelerometers
to remove effects from
non-gravitational forces and Global Positioning Systems to provide both the
precise time-tags for the recorded data and the positions of
the satellites over Earth.

To continue the critical Earth gravity variation data recorded by
GRACE, NASA has scheduled the launch of the GRACE
Follow On mission to August 2017.
 The GRACE FO mission would re-fly the identical GRACE
spacecraft and instruments, but supplement the micrometre-level
accuracy microwave measurement with a laser interferometer of
nanometre-level accuracy\footnote{Please see
\url{https://www.aei.mpg.de/18528/04_Grace_Follow-on}}.
From the detailed simulations of the laser ranging system of GRACE FO
\cite{Pierce2008,Loomis2012,Sheard2012}, in the signal band
$10^{-2}Hz\sim10^{-1}Hz$, the range measurements accuracy is about
$100nm/\sqrt{Hz}$ for $270km$
satellites separation and $1pm/\sqrt{Hz}\sim0.1nm/\sqrt{Hz}$ for $50km$
separation. Near
the orbital frequency $\sim10^{-4}Hz$, the accuracy is still about
$200\mu m/\sqrt{Hz}$ for $270km$ satellites separation and $10\mu m/\sqrt{Hz}$
for $50km$
separation.

The GRACE and GRACE FO missions, in the first place, are not designed
for the tests of relativistic theories of gravitation. While, as discussed
above, the great accuracy in the range and range rate measurements, especially
for the case of the laser ranging system
of GRACE FO, are in principle possible to place strong constraints on
certain relativistic gravitational theories, please also consult the
works \cite{Iorio2012,Iorio2012a,Iorio2012b} along this line.

\section{Non-dynamical Chern-Simons modified gravity}\label{sec:NCS}

We give a brief introduction to the non-dynamical formulation of CS
modified gravity, for detailed discussions please consult
\cite{Alexander2009,Alexander2007,Alexander2007a}.
The geometric units $c=G=1$ are adopted. The action for the
non-dynamical CS gravity reads
\[
S:=S_{GR}+S_{CS}+S_{matt},
\]
where
\begin{eqnarray}
S_{GR} & = & \frac{1}{16\pi}\int d^{4}x\sqrt{-g}R,\nonumber\\
S_{CS} & = & \frac{\alpha}{4}\int d^{4}x\sqrt{-g}\theta
R^{\star}R,\label{CSaction}
\end{eqnarray}
and $S_{matt}$ is the action from the matter fields which is independent
of $\theta$. $g$ is the determinant of the metric and the Pontryagin
density reads
\begin{equation}
R\;^{\star}R=\frac{1}{2}\epsilon^{cdef}R_{\; bef}^{a}R_{\;
acd}^{b}.\label{eq:pon}
\end{equation}
The magnitude of the CS extension is controlled by the coupling field
$\theta$, which is externally prescribed and depends
on the mass scale of the specific theory that under the consideration.
$\theta$ can also be
viewed as the deformation function, and the difference between CS
gravity and GR is proportional to the deformation parameters $\nabla_{a}\theta$
and $\nabla_{a}\nabla_{b}\theta$. In this work, the most popular choice, called
the canonical coupling \cite{Jackiw2003}, is adopted, where $\theta$ is a
spatially homogeneous function and depends linearly on time. Therefore the
deformation parameter contains only $\dot{\theta}$. With such choice,
spacetime-dependent reparameterization of spacial variables and time
translation remain symmetries of the CS modified theory
\cite{Alexander2007,Alexander2009}.

The field equation of the non-dynamical CS gravity is obtained by
varying the action with respect to the metric
\begin{equation}
R_{ab}-\frac{1}{2}g_{ab}R+16\pi\alpha C_{ab}=8\pi T_{ab},\label{eq:field_eq}
\end{equation}
where $C_{ab}$ is the 4-dimensional generalization of the Cotton-York
tensor
\begin{equation}
C^{ab}=\nabla_{c}\theta\epsilon^{cde(a}\nabla_{e}R_{\:\:
d}^{b)}+\frac{1}{2}\nabla_{c}\nabla_{d}\theta\epsilon^{efd(a}R_{\;\;\;\:\:
fe}^{b)c}.\label{eq:cotton}
\end{equation}
The introduction of the new scalar degree of freedom $\theta$ also
gives rise to the new constraint
\begin{equation}
\nabla_{a}C^{ab}=-\frac{1}{8}\nabla^{b}\theta(^{\star}RR)=0.\label{eq:constraint
}
\end{equation}
If the above constraint is satisfied, from eq.(\ref{eq:field_eq}),
the Bianchi identities and the equations of motion for matter fields
$\nabla_{a}T^{ab}=0$
are recovered, which rank the non-dynamical
CS gravity a metric theory.

In the weak field and slow motion limits, the Parametrized Post-Newtonian
(PPN) metric of the non-dynamical CS gravity outside a compact source
was carefully worked out in \cite{Alexander2007,Alexander2007a}.
As mentioned before, the non-dynamical CS gravity differs from GR
only in the gravitomagnetic sector
\begin{equation}
g_{0i}^{CS}=g_{0i}^{GR}+\chi(r\nabla\times\mathbf{V})_{i},\label{eq:CSGM}
\end{equation}
here $r$ denotes the distance to the mass center of the compact source
and $V_{i}$ is the PN potential, see \ref{sec:The-Standard-PPN}.
The dimensionless parameter $\chi=\frac{32\pi\alpha\dot{\theta}}{r}$
is the new PN parameter for non-dynamical CS gravity, and the CS mass
scale reads \cite{Alexander2007,Alexander2007a}
\begin{equation}
Mcs=\frac{1}{8\pi\alpha\dot{\theta}}=\frac{4}{\chi r}.\label{eq:mass}
\end{equation}

\section{The range observable of non-dynamical Chern-Simons gravity
\label{sec:CS}}

\subsection{The basic settings}
According to the SST missions introduced in
section  \ref{sec:GRACE}, we study the range observable between the two
satellites, which are modeled here as two proof masses orbiting Earth one after
another along
nearly circular orbits.
We restrict ourselves to the so-called ``semi-conservative'' metric
theories, which are based on action principles and respect the conservation
law of 4-momentum \cite{Will2014}. Therefore, the relevant PN
parameters are
$\{\gamma,\ \beta,\ \xi,\ \alpha_{1},\ \alpha_{2}\}$
together with the additional CS parameter $\chi$, please see \cite{Will2014}
or \ref{sec:The-Standard-PPN} for the parametrized PN formalism. The
PN coordinates system $\{t,x^{i}\}$ outside Earth is chosen as follows.
The mass center of Earth is set at the origin. The basis
 $(\frac{\partial}{\partial x^{3}})^{a}$
is set to parallel to the direction of the Earth angular momentum
$\mathbf{J}$, $(\frac{\partial}{\partial x^{1}})^{a}$ is pointing
to a reference star $\Upsilon$ and $(\frac{\partial}{\partial x^{2}})^{a}$
determined by the right-hand rule $(\frac{\partial}{\partial
  x^{1}})^{a}\times(\frac{\partial}{\partial x^{2}})^{a}
=(\frac{\partial}{\partial x^{3}})^{a}$, see Fig.\ref{fig:coor} for illustration.
Such coordinate directions are tied to the remote stars, and the time
$t$ is measured by the observers at asymptotically flat region. Within
our coordinate system the PN metric outside Earth reads
\begin{eqnarray*}
g_{00} & = & -1+2U-2\beta U^{2}-2\xi\Phi_{W}+(2\gamma+2-2\xi)\Phi_{1}\\
 &  & +2(3\gamma-2\beta+1+\xi)\Phi_{2}+2\Phi_{3}+2(3\gamma-2\xi)\Phi_{4}\\
 &  & +2\xi\mathcal{A}-(\alpha_{1}-\alpha_{2})w^{2}U-\alpha_{2}w^{i}w^{j}U_{ij}-2\alpha_{1}w^{i}V_{i}+\mathcal{O}(\epsilon^{6}),\\
g_{0i} & = & -\frac{1}{2}(4\gamma+3+\alpha_{1}-\alpha_{2}-2\xi)V_{i}-\frac{1}{2}(1+\alpha_{2}+2\xi)W_{i}\\
 &  & +\chi r(\nabla\times\mathbf{V})_{i}-\frac{1}{2}(\alpha_{1}-2\alpha_{2})w_{i}U-\alpha_{2}w^{j}U_{ij}+\mathcal{O}(\epsilon^{5}),\\
g_{ij} & = & (1+2\gamma U)\delta_{ij}+\mathcal{O}(\epsilon^{4}),
\end{eqnarray*}
please see \ref{sec:The-Standard-PPN} for the PN potentials.
For low and medium Earth orbits experiments, the magnitude of $\epsilon$
is about $10^{-5}.$

We model Earth as an ideal and uniform rotating spherical body.
The preferred-frame and the preferred-location effects are tightly
constrained by observations, and we now have the upper bounds of the
related PN parameters as $\alpha_{1}\sim4\times10^{-5},\
\alpha_{2}\sim2\times10^{-9},
\ \alpha_{3}\sim4\times10^{-20}$
and $\xi\sim10^{-9}$, please see Tab.\ref{tab: PPN value} or \cite{Will2014}
for more details. Generally, the coordinate velocity $w$ of the PPN
coordinate system relative to the mean rest-frame of the universe
is believed to be small, that $w\sim\mathcal{O}(\epsilon)$
\cite{Will1972,Will1993,Will2014}. Therefore, the gradients produced
by the preferred-frame and the preferred-location effects between the two orbiting
satellites will be smaller than $10^{-21}s^{-2}$, which will produce a
relative acceleration smaller than $2\times10^{-16}m/s^2$. This is too small to
be seen by the present day and future planned SST missions and will be
ignored in this work. The above metric can then be cast into a rather simple form
\begin{eqnarray}
 &  & g_{\mu\nu}=\nonumber \\
 &  & \left(\begin{array}{cccc}
-1+\frac{2M}{r}-\frac{2\beta M^{2}}{r^{2}} & (\frac{\Delta x^{2}}{r^{3}}+\frac{3\chi x^{1}x^{3}}{2r^{4}})J & (-\frac{\Delta x^{1}}{r^{3}}+\frac{3\chi x^{2}x^{3}}{2r^{4}})J & -\frac{\chi[(x^{1})^{2}+(x^{2})^{2}-2(x^{3})^{2}]}{2r^{4}}J\\
\\
(\frac{\Delta x^{2}}{r^{3}}+\frac{3\chi x^{1}x^{3}}{2r^{4}})J & 1+\frac{2\gamma M}{r} & 0 & 0\\
\\
(-\frac{\Delta x^{1}}{r^{3}}+\frac{3\chi x^{2}x^{3}}{2r^{4}})J & 0 & 1+\frac{2\gamma M}{r} & 0\\
\\
-\frac{\chi[(x^{1})^{2}+(x^{2})^{2}-2(x^{3})^{2}]}{2r^{4}}J & 0 & 0 & 1+\frac{2\gamma M}{r}
\end{array}\right),\nonumber \\
\label{eq:metric}
\end{eqnarray}
where $r=\sqrt{\delta_{ij}x^{i}x^{j}}$, $\Delta=1+\gamma+\frac{1}{4}\alpha_{1}$,
and $M,\ \mathbf{J}$ are the asymptotically measured total mass and
angular momentum of Earth
\begin{eqnarray*}
M & = & \int\rho[1+(\gamma+1)v^{2}+(3\gamma-2\beta+1)U+\frac{\Pi}{\rho}+3\gamma\frac{p}{\rho}]d^{3}x,\\
\mathbf{J} & = & \int\rho(\mathbf{x}\times\mathbf{v})d^{3}x.
\end{eqnarray*}
For a satellite orbiting Earth with velocity $v,$ one has the basic
order relations
\begin{eqnarray}
v^{2}\sim\frac{M}{r}\sim\mathcal{O}(\epsilon^{2}),\quad\: v^{4}\sim\frac{M^{2}}{r^{2}}\sim\frac{Jv}{r^{2}}\sim\mathcal{O}(\epsilon^{4}).\label{eq:MJorder}
\end{eqnarray}

One should notice that the leading gradients acting on the two satellites
are of the Newtonian ones, the PN
gradients are generally of $\mathcal{O}(\epsilon^2)\sim 10^{-10}$ times smaller
than the Newtonian
gradients. At the PN level, the deviations of the centered matter source from
ideal uniform sphere will give rise to multipolar corrections to the Newtonian
potential
in the $g_{00}$ component
\begin{equation}
U=\frac{M}{r}+\frac{M}{r}\sum_{l=2}^{\infty}\frac{R^{l}}{r^{l}}\sum_{m=-l}^{l}C_
{lm}Y_{lm},\label{eq:multipole}
\end{equation}
where $R$ is the mean radius of Earth and $C_{lm}$ is the coefficient of the
spherical harmonic component. Potentials from these multiples belongs to
purely non-relativistic effects, and along low Earth orbits they are
generally smaller
than $\mathcal{O}(\epsilon^{3})$ with the only exception of the
$J_{2}\sim10^{-3}\frac{R^{2}}{r^{2}}\frac{M}{r}$
component\footnote{\url{
http://op.gfz-potsdam.de/grace/results/grav/g002_eigen-grace02s.html}}. We
ignore such multiples in the following theoretical
analysis in this section. The effects of these multiples on the measurements
and the possible data analysis methods will be briefly discussed in section
\ref{sec:measurements}.

\subsection{The geodesic deviation equation}

The first step studies of the geodesic motions of proof masses in CS
modified gravity and their orbital observable can be found in
\cite{Smith2008,Chen2010}.
While, in this work, since we are interested in the range observable
between the two satellites instead,
the geodesic deviation equation that describes their relative motions
will be a proper starting point
\begin{equation}
T^{b}\nabla_{b}T^{c}\nabla_{c}X^{a}+R_{bcd}^{\ \ \:\:\:\:\:
a}T^{b}T^{d}X^{c}=0.\label{eq:deviation}
\end{equation}
Here $T^{a}$ denotes the 4-velocity of the reference satellite and
$X^{a}$ the connection vector pointing from the reference satellite to the
second one. We then introduce the tetrad
$\{(\mathbf{e}_{I})^{a},I=0,1,2,3\}$
carried by the reference satellite with $(\mathbf{e}_0)^a=T^a$. Through such
local tetrad, we can map the above
equation into the local frame of the reference satellite
\begin{eqnarray}
\frac{d^{2}}{d\tau^{2}}X^{I} &
=&-2\gamma^I_{\:\:\:\:J0}\frac{d}{d\tau}X^J
-(\frac{d}{d\tau}
\gamma^I_{\:\:\:\:J0}
+\gamma^K_{\:\:\:\:J0}\gamma^I_{\:\:\:\:K0})X^J\nonumber{}\\
&&-K_J^{\:\:I}X^J.\label{eq:localdev}
\end{eqnarray}
Here, $\tau$ is the proper time measured by the reference satellite, and
$\gamma^I_{\:\:\:\:JK}=(\mathbf{e}^I)^{\nu}(\mathbf{e}_J)^{\mu}\nabla_{\mu}
(\mathbf{e}_K)_{\nu}$
are the Ricci rotation coefficients. According to the convention,
the upper-latins $\{I,\ J,\ K,...\}$ are used to index tensor components
under the local tetrad $(\mathbf{e}_{I})^{a}$. $(\mathbf{e}_I)^{\mu}$
can be viewed as the transformation
matrix from local system to the Earth centered PN system,
and $(\mathbf{e}^I)_{\mu}$ the inverse.
The first lines of the right hand side of the
above equation come from the gradients of inertial forces, which are
resulted from the relative rotation of the local frame to the Fermi
shifted frame. The last line comes from the tidal forces from
spacetime curvature,
where the the tidal matrix is defined as
\begin{equation}
K_{J}^{\ \ I}=R_{\lambda\nu\rho}^{\ \ \ \ \ \ \mu}T^{\lambda}T^{\rho}(\mathbf{e}_J)^{\nu}(\mathbf{e}^I)_{\mu}.\label{eq:gradient}
\end{equation}

\subsection{The order estimations of the geodesic deviations}

\begin{figure}
\center
\includegraphics[scale=0.68]{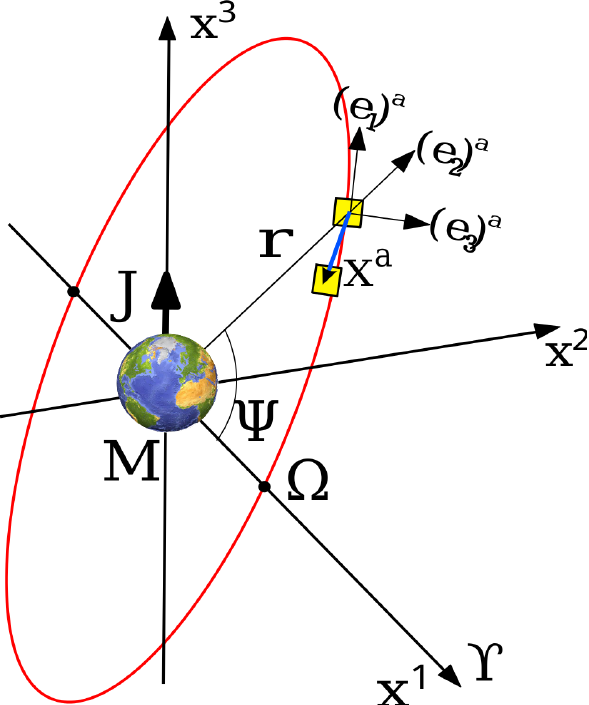}
\caption{The settings of the PN reference frame and the local frame of the
reference satellite. The orbits under consideration are circular orbits with
the longitude of ascending node $\Omega=0$. As illustrated in the figure, the
local tetrad carried by the reference satellite is defined as follows,
$(\mathbf{e}_{1})^{a}$
is set along the direction of motion of the reference satellite,
$(\mathbf{e}_{2})^{a}$
along the radial direction and
$(\mathbf{e}_{3})^{a}=(\mathbf{e}_{1})^{a}\times(\mathbf{e}_{2})^{a}$.}
\label{fig:coor}
\end{figure}

Eq.(\ref{eq:localdev}) is a system of ordinary differential equations
that evaluated along the orbits of the reference satellite in its local
frame. From dimensional analysis, up to the required 1PN level,
eq.(\ref{eq:deviation}) or eq.(\ref{eq:localdev}) will have the
following form
\begin{equation}
\frac{d^{2}X}{d\tau^{2}}\sim\frac{1}{r}\frac{|X|}{r}
(\mathcal{O}(\epsilon^{2})+\mathcal{O}(\epsilon^{4})+... ) .\label{eq:orderdev}
\end{equation}
To clearly book-keep all the possible perturbation terms which
appear in eq.(\ref{eq:localdev}) and to understand the physical picture, we
take the following approach.
We first ignore the rotation of Earth, and the metric now
reduce to the 1PN approximation of a spherical symmetric spacetime
\begin{equation}
g_{\mu\nu}^{\mathbf{S}}=\left(\begin{array}{cccc}
-1+\frac{2M}{r}-\frac{2\beta M^{2}}{r^{2}} & 0 & 0 & 0\\
0 & 1+\frac{2\gamma M}{r} & 0 & 0\\
0 & 0 & 1+\frac{2\gamma M}{r} & 0\\
0 & 0 & 0 & 1+\frac{2\gamma M}{r}
\end{array}\right).\label{eq:spherical}
\end{equation}
For proof mass orbiting around the centered source in the above
spherical symmetric spacetime, one can work out the equation of motion
from the geodesic equation as
\cite{Petit2010}
\begin{eqnarray}
 \frac{d^2\mathbf{x}}{dt^2}=-\frac{M}{r^3}\mathbf{x}+\frac{M}{r^3}
\left((\frac{2(\gamma+\beta)M}{r}-\gamma
v^2)\mathbf{x}+2(1+\gamma)(\mathbf{x}\cdot\mathbf{v})\mathbf{v}
\right)+\frac{1}{r}\mathcal{O}(\epsilon^6).
\end{eqnarray}
At the 1PN level, circular orbits exist with the orbital frequency
$\omega$
\begin{equation}
\omega^{2}=\frac{1}{a^{2}}[\frac{M}{a}+(3-\gamma-2\beta)\frac{M^{2}}{a^{2}}
+\mathcal{O}(\epsilon^{6})].\label{eq:omega}
\end{equation}
Therefore, for the ideal case, we set the two satellites to follow, one after
another, the same circular orbit
\begin{equation}
x^{1}=a\cos\Psi,\ x^{2}=a\cos i\sin\Psi,\ x^{3}=a\sin i\sin\Psi,\label{eq:orbit}
\end{equation}
where the longitude of ascending node $\Omega$ is set to be zero
and $\Psi=\omega\tau$ denotes the orbital
phase, see figure.\ref{fig:coor}. Then, the most natural choice of the
local tetrad for the evaluations of the range observable is the followings, that
we set $(\mathbf{e}_{1})^{a}$ along the direction of motion of the
reference satellite, $(\mathbf{e}_{2})^{a}$ along the radial direction,
$(\mathbf{e}_{3})^{a}=(\mathbf{e}_{1})^{a}\times(\mathbf{e}_{2})^{a}$
determined by the right hand rule
and $(\mathbf{e}_{0})^{a}=T^{a}$, see again figure.\ref{fig:coor}.
Under such tetrad, the second satellite will stay static and the component
of the connection vector will have the simple form
$X^I=\{0,\rho_0,0,0\}$,
here $\rho_0$ measures the separation or the range between the two satellites.
Such results in the local frame is due to the
cancellations between the gradients of
centripetal and centrifugal forces in this Earth pointing system.

Now, we ``turn on'' the Earth rotation and recover the $g_{0i}$
components proportional to the source angular momentum $J$, which
can be taken as the 1PN perturbations act on the above circular orbits.
The two satellites system now begins to ``feel'' the tidal force from the
gravitomagnetic sector. Such gravitomagnetic tidal force will drive
the second satellite to deviate from its original position relative to the
reference
one, which means that the components of the connection vector will now
change with time
\[
X^I(\tau)=\{0,\rho_0+\delta^{1}(\tau),\delta^{2}(\tau),\delta^{3}
(\tau)\}.
\]
Here, we need to estimate the magnitudes of this new family of small
quantities, that the small deviations $\{\delta^{i}(\tau)\}$
driven by the gravitomagnetic tidal force, their time derivatives
$\{\dot{\delta^{i}}(\tau)\}$
and the second time derivatives $\{\ddot{\delta^{i}}(\tau)\}$. From
dimensional analysis, in the reference satellite local frame, the magnitude
of the periodic gravitomagnetic force acting on the second satellite is
\begin{equation}
\ddot{\delta^{i}}\sim\frac{Jv\rho_0}{a^{4}}\sim\frac{1}{a}\frac{\rho_0}{a}
\mathcal{O}(\epsilon^{4}),\label{eq:neworder3}
\end{equation}
and its frequency is of the orbital frequency
$\omega\sim\sqrt{M/a^{3}}$.
Thus one has the important order relations
\begin{eqnarray}
\dot{\delta}^{i} & \sim &
\frac{Jv\rho_0}{a^{4}\omega}\sim\frac{\rho_0}{a}\mathcal{O}(\epsilon^{3}),
\label{eq:neworder1}\\
\delta^{i} & \sim & \frac{Jv\rho_0}{a^{4}\omega^{2}}\sim
\rho_0\mathcal{O}(\epsilon^{2}).\label{eq:neworder2}
\end{eqnarray}
Therefore, with the above analysis and eq.(\ref{eq:orderdev}), up
the to the 1PN level the geodesic deviation equation eq.(\ref{eq:localdev})
will only contain the following terms
\begin{equation}
\ddot{\delta}=\underbrace{\frac{1}{a}\frac{\rho_0}{a}\mathcal{O}(\epsilon^{2})}_
{Newtonian\
part}+\underbrace{\frac{1}{a}\dot{\delta}\mathcal{O}(\epsilon)+\frac{1}{a^2}
\delta\mathcal{O}(\epsilon^{2})+\frac{1}{a}\frac{\rho_0}{a}\mathcal{O}(\epsilon^
{4})}_{1PN\ part}.\label{eq:orderDev}
\end{equation}

At last, we work out the relation between the range variations $\delta\rho$
and the variations of the components of the connection vector. For
nearly circular orbits,
the relations between the components of the connection vector
$X^I$ and the coordinates $x^{I}_2$
of the second satellite under the local frame is
\[
x^{I}_2=X^{I}+\mathcal{O}(\frac{|X|^{2}}{a}).
\]
Then the relation between the variations of the coordinates of the second
satellite $\delta x^{I}_2$ and the variations of the connection
vector $\delta^{i}$ reads
\[
\delta x^{I}_2(\tau)=\delta^{I}(\tau)+a\frac{\rho_0^{2}}{a^{2}}\mathcal{O}
(\epsilon^ { 2 })=\delta^{I}(\tau)(1+\mathcal{O}(\frac{\rho_0}{a})+...).
\]
For GRACE and GRACE Follow On missions, the choice of $\rho_0$ ranges
from $50km$ to $270km$, and therefore $\frac{\rho_0}{a}\leq3\times10^{-2}$.
Thus, to the leading order, the range
variation will have the simple from
\begin{eqnarray}
\delta\rho(\tau)
&=&\delta^1(\tau)+\sum_{i=2,3}\mathcal{O}(\frac{\rho_0}{a})\delta^{i}(\tau)+
[ \frac { (\delta^2(\tau))^2 } {
2\rho_0^2}+\frac{(\delta^3(\tau))^2}{2\rho_0^2}]
\rho_0\nonumber\\
&=&\delta^1(\tau)+
\sum_{i=2,3}\delta^i(\tau)[\mathcal{O}(\frac{
\rho_0}{a})+\mathcal{O}(\epsilon^2)].\label{eq:rho2delta}
\end{eqnarray}

\subsection{The equation of motion in the local frame}

As discussed in the last subsection, the geodesic deviation of the second
satellite relative to the reference one along the nearly circular orbit is
produced
by the 1PN gravitomagnetic tidal force, which are of terms
proportional to $\frac{\rho_0}{a^2}\mathcal{O}(\frac{Jv}{a^{4}})\sim
\frac{\rho_0}{a^2}\mathcal{O}(\epsilon^{4})$. Therefore, from
eq.(\ref{eq:orderDev}), to calculate such 1PN forces, one only need to work
with the Newtonian (Keplerian) orbits. From the orbital choices of GRACE
and GRACE FO missions, we work
out here the range observable of non-dynamical CS gravity for the case
of circular orbits, that
eq.(\ref{eq:orbit}). The effect of small orbital eccentricities
$e\sim2\times10^{-3}$ will be
left in future works concerning real data analysis.

Along the circular orbit, eq.(\ref{eq:orbit}),
the 4-velocity of the reference satellite reads
\begin{eqnarray}
T^{a} & =&\frac{dt}{d\tau}(\frac{\partial}{\partial
t})^{a}+a\omega[-\sin\Psi(\frac{\partial}{\partial x^{1}})^{a}
 \nonumber\\
  && +\cos i\cos\Psi(\frac{\partial}{\partial x^{2}})^{a}+\sin
i\cos\Psi(\frac{\partial}{\partial x^{3}})^{a}].\label{eq:Z}
\end{eqnarray}
The ratio $\frac{dt}{d\tau}$ can be derived from the line element
$d\tau^{2}=-g_{\mu\nu}dx^{\mu}dx^{\nu}$ evaluated along the orbit
\begin{eqnarray}
\frac{dt}{d\tau} & = &
1+\frac{a^{2}\omega^{2}}{2}+\frac{M}{a}+\mathcal{O}(\epsilon^4).\label{eq:time}
\end{eqnarray}
For the tetrad attached to the reference satellite defined in the last
subsection,
we first set $(\mathbf{e}_{0})^{a}=T^{a}$, and following
the Gram-Schmidt process the three spacial bases can be worked out
as
\begin{eqnarray*}
 (\mathbf{e}_{1})^{a} & =&  a\omega(\frac{\partial}{\partial t})^{a}
+(1+\frac{a^{2}\omega^{2}}{2}-\frac{\gamma
M}{a})[-\sin\Psi(\frac{\partial}{\partial x^{1}})^{a}\\
 && +\cos i\cos\Psi(\frac{\partial}{\partial x^{2}})^{a}+\sin
i\cos\Psi(\frac{\partial}{\partial x^{3}})^{a}],\\
(\mathbf{e}_{2})^{a} & = & (1-\frac{\gamma
M}{a})[\cos\Psi(\frac{\partial}{\partial x^{1}})^{a} +\cos
i\sin\Psi(\frac{\partial}{\partial x^{2}})^{a}\\&&+\sin
i\sin\Psi(\frac{\partial}{\partial x^{3}})^{a}],\\
(\mathbf{e}_{3})^{a} & = &  (1-\frac{\gamma
M}{a})[\cos i(\frac{\partial}{\partial
x^{3}})^{a}-\sin i(\frac{\partial}{\partial x^{2}})^{a}].
\end{eqnarray*}
The transformation matrices then read

\begin{eqnarray}
&&(\mathbf{e}_I)^{\mu}  =\nonumber\\&&
\begin{small}
\left(\begin{array}{cccc}
1+\frac{a^{2}\omega^{2}}{2}+\frac{M}{a} & -a\omega\sin\Psi & a\omega\cos
i\cos\Psi & a\omega\sin i\cos\Psi\\
a\omega & -(1+\frac{a^{2}\omega^{2}}{2}-\frac{\gamma M}{a})\sin\Psi &
(1+\frac{a^{2}\omega^{2}}{2}-\frac{\gamma M}{a})\cos i\cos\Psi &
(1+\frac{a^{2}\omega^{2}}{2}-\frac{\gamma M}{a})\sin i\sin\Psi\\
0 & \left(1-\frac{\gamma M}{a}\right)\cos\Psi & \left(1-\frac{\gamma
M}{a}\right)\cos i\sin\Psi & \left(1-\frac{\gamma M}{a}\right)\sin i\sin\Psi\\
0 & 0 & -\left(1-\frac{\gamma M}{a}\right)\sin i & \left(1-\frac{\gamma
M}{a}\right)\cos i
\end{array}\right),\end{small}\nonumber\\
\label{eq:e}
\end{eqnarray}
\begin{eqnarray}
&&(\mathbf{e}^I)_{\mu}  =\nonumber\\&&\left(\begin{array}{cccc}
1+\frac{a^{2}\omega^{2}}{2}-\frac{M}{a} & -a\omega & 0 & 0\\
a\omega\sin\Psi & -(1+\frac{a^{2}\omega^{2}}{2}+\frac{\gamma M}{a})\sin\Psi &
(1+\frac{\gamma M}{a})\cos\Psi & 0\\
-a\omega\cos i\cos\Psi & (1+\frac{a^{2}\omega^{2}}{2}+\frac{\gamma M}{a})
\cos i\cos\Psi & (1+\frac{\gamma M}{a})\cos i\sin\Psi &
-(1+\frac{\gamma M}{a})\sin i\\
-a\omega\sin i\cos\Psi & (1+\frac{a^{2}\omega^{2}}{2}+\frac{\gamma M}{a})
\sin i\cos\Psi & (1+\frac{\gamma M}{a})\sin i\sin\Psi &
(1+\frac{\gamma M}{a})\cos i
\end{array}\right).\nonumber\\
\label{eq:ie}
\end{eqnarray}

The explicit forms of the Christofell symbols
$\Gamma^{\lambda}_{\:\:\ \mu\nu}$ and the tidal matrix from the Riemann curvature
$K_{J}^{\ \ J}=R_{\lambda\nu\rho}^{\ \ \ \ \ \
\mu}T^{\lambda}T^{\rho}(\mathbf{e}_J)^{\nu}(\mathbf{e}^I)_{\mu}$
are worked out up to 1PN level in
\ref{sec:The-Christoffel-Symbol}.
Now, with all the results gathered here, we substitute the connection
vector $X^I$, the transformation matrices eq.(\ref{eq:e}),
eq.(\ref{eq:ie}), the Christoffel symbols eq.(\ref{eq:G00m})-eq.(\ref{eq:G0ij})
and the curvature tidal matrices eq.(\ref{eq:TN})-eq.(\ref{eq:TCS}) into the
geodesic deviation equation eq.(\ref{eq:localdev}).
After the heavy works of simplifications and, according to
eq.(\ref{eq:rho2delta}), ignoring all the terms beyond
$\frac{1}{a^{2}}\delta^{i}\mathcal{O}(\epsilon^{2})$,
$\frac{1}{a}\dot{\delta}^{i}\mathcal{O}(\epsilon)$
and $\frac{\rho_0}{a^{2}}\mathcal{O}(\epsilon^{4})$
we have the rather simple forms of the equations of motions
that govern the deviations of the second satellite
 \begin{eqnarray}
 \ddot{\delta}^{1}(\tau)&=&-2\omega\dot{\delta}^{2}(\tau)-\chi\frac{
 \rho_0J\omega\sin i\cos(\omega\tau)}{2a^{3}} ,\label{eq:inline}\\
 \ddot{\delta}^{2}(\tau)&=&2\omega\dot{\delta}^{1}(\tau)+3\omega^{2}\delta^{2}
 (\tau)+\chi\frac{\rho_0J\omega\sin
   i\sin(\omega\tau)}{2a^{3}}.\label{eq:delta2}\\
 \ddot{\delta}^{3}(\tau)&=&-\omega^2\delta^3(\tau)
 +2\Delta\frac{\rho_0J\omega\sin
   i\cos(\omega\tau)}{a^3}
 .\label{eq:delta3}
\end{eqnarray}
As one should expect that the deviation $\delta^3(\tau)$ in the
direction perpendicular to the
$(\mathbf{e}_1)^a-(\mathbf{e}_2)^a$ plane will not couple into
the first two equations. This is because
the 1PN deviation $\delta^3(\tau)$ is perpendicular to both the along-track
and radial directions, and can only alter the
range $\rho(\tau)$ and the semi-major $a$ of the second satellite at the
2PN level.
On the other hand, within the orbital plane, the deviation $\delta^2(\tau)$
along the radial direction
does couple to the deviation $\delta^1(\tau)$ in the along-track direction and
vise versa through
the Coriolis effect.

An important feature of these geodesic deviation equations is that the 1PN
deviations in the along-track and radial directions depend only on
the tidal forces from the CS extension term in eq.(\ref{CSaction}). The 1PN
tidal forces proportional to the standard PN parameters $\{\gamma,\:\beta,
\:\alpha_1,\alpha_2\}$ do not appear in the motions along these two
directions. This is due to the fact that the gravitomagnetic gradients in
standard
parity-preserving metric theories evaluated along circular orbits
will only affect the deviations in the transverse direction, which, as
discussed above, will only
affect the deviations within the orbital plane at 2PN
level. Therefore, being a true advantage, the range
variations at the orbital frequency can be used to distinguish CS gravity from
standard parity-preserving metric theories (including GR).

At last, one should notice that the simple forms of the geodesics deviation
equations under the local frame, that
eq.(\ref{eq:inline})-eq.(\ref{eq:delta3}), and the decoupling of the motions in
the transverse direction from the along-track motions stay true only
when the deviations are within the 1PN level, that
$\delta^i(\tau)<\rho_0\mathcal{O}(\epsilon)$ and $\dot{\delta}^i(\tau)
<\frac{\rho_0}{a}\mathcal{O}(\epsilon^2)$.
When the deviations
$\delta^i(\tau)$, $\dot{\delta}^i(\tau)$
are beyond the 1PN level, terms of
$\frac{1}{a^2}\delta^i\mathcal{O}(\epsilon^4)$ and
$\frac{1}{a}\dot{\delta}^i\mathcal{O}(\epsilon^3)$ that are ignored
at the first place will begin to play important roles in determining the
relative motions. Also, the gravitomagnetic perturbations of the orbit of
the reference satellite must be included into the geodesic
deviation equations. Thus, for large deviations
$\delta^i\geq\rho_0\mathcal{O}(\epsilon)$ and $\dot{\delta}^i(\tau)
\geq \frac{\rho_0}{a}\mathcal{O}(\epsilon^2)$,
eq.(\ref{eq:inline})-eq.(\ref{eq:delta3}) will break down, and
the full geodesic deviation equations will turn out to be
very complicate and can hardly be solved analytically.

\subsection{The range observable}
\begin{figure*}
\center
\includegraphics[scale=0.6]{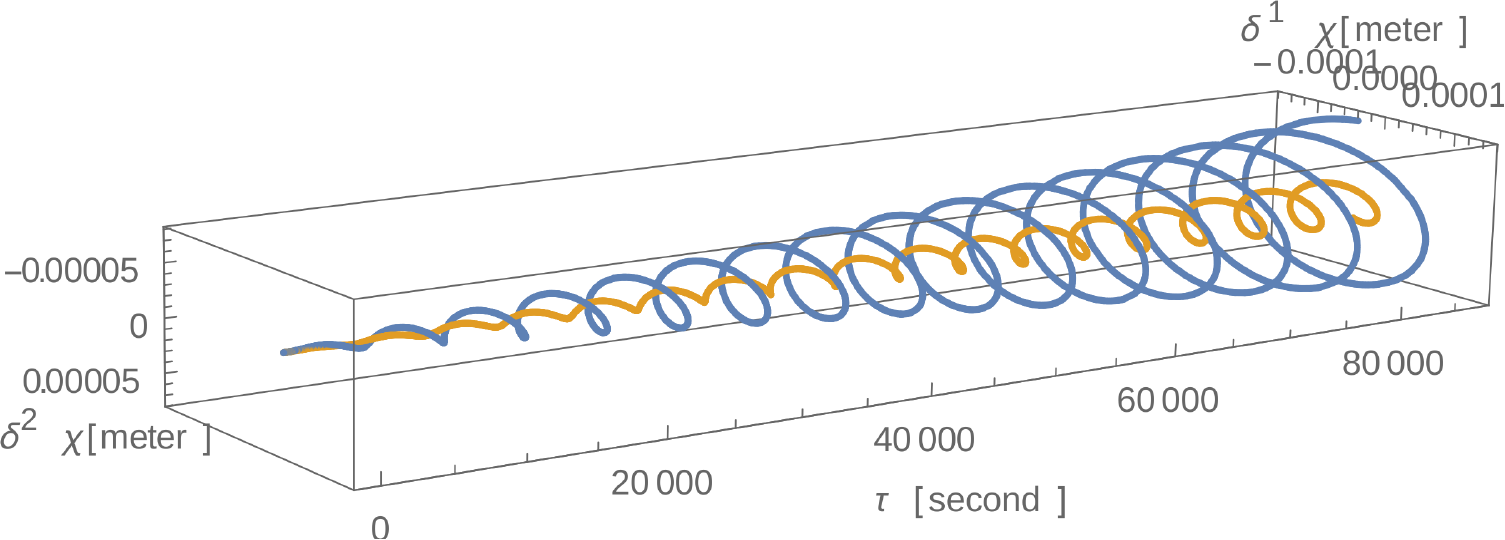}
\caption{The relative motions of the second satellite to the
  reference one in the $(\tilde{\mathbf{e}}_1)^a-(\tilde{\mathbf{e}}_2)^a$
  plane. The orbits are chosen as near-circular polar orbits with altitudes
  as $460km/250km$. The satellites separations are chosen as $220km/50km$.}
\label{fig:planemotion}
\end{figure*}

\begin{figure*}
\center
\includegraphics[scale=0.55]{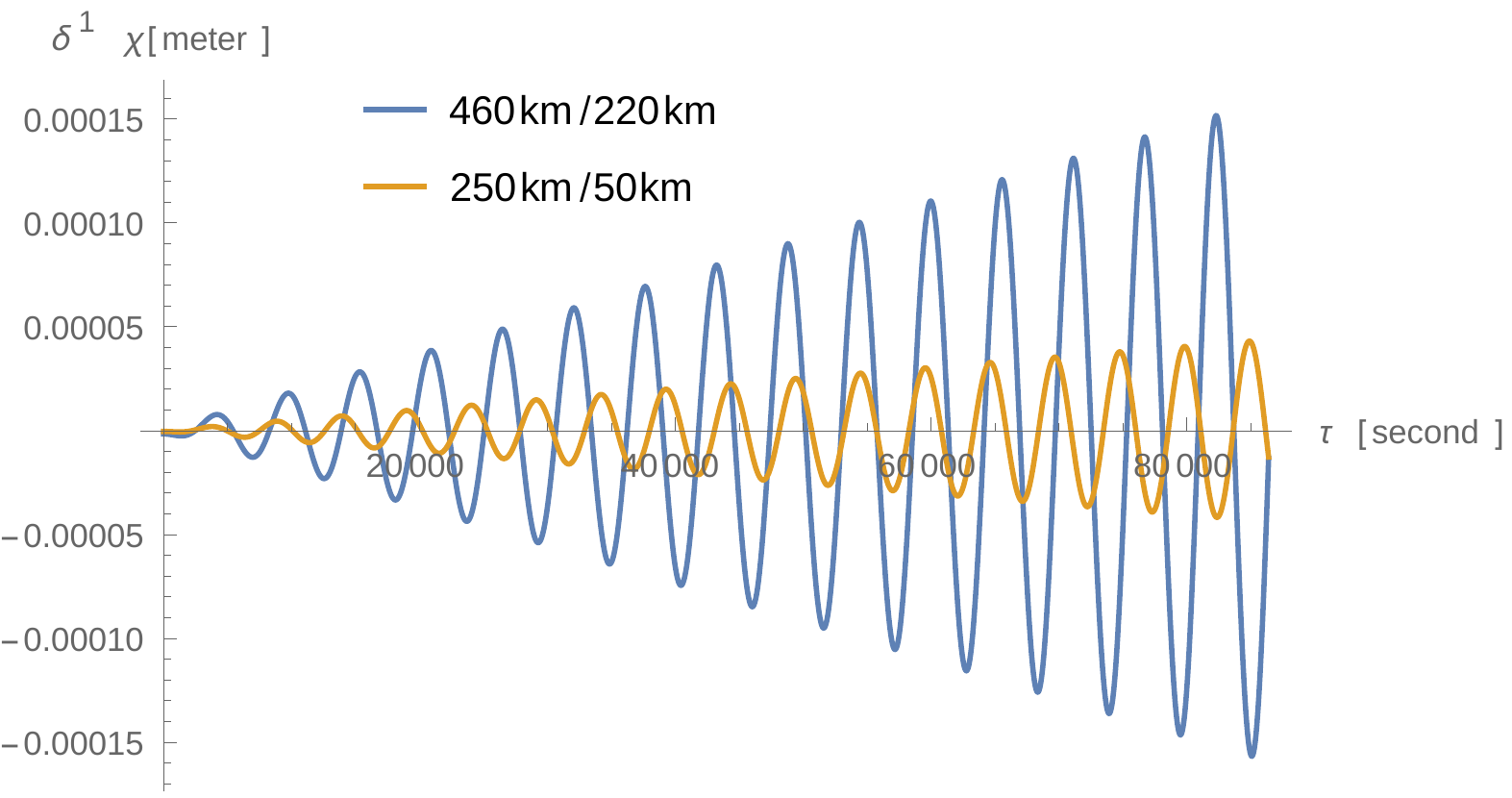}
\caption{The range variations in the along-track direction between the two
satellites. The orbits are chosen as near-circular polar orbits with altitudes
  as $460km/250km$. The satellites separations are chosen as $220km/50km$.}
\label{fig:delta1}
\end{figure*}

\begin{figure*}
\center
\includegraphics[scale=0.55]{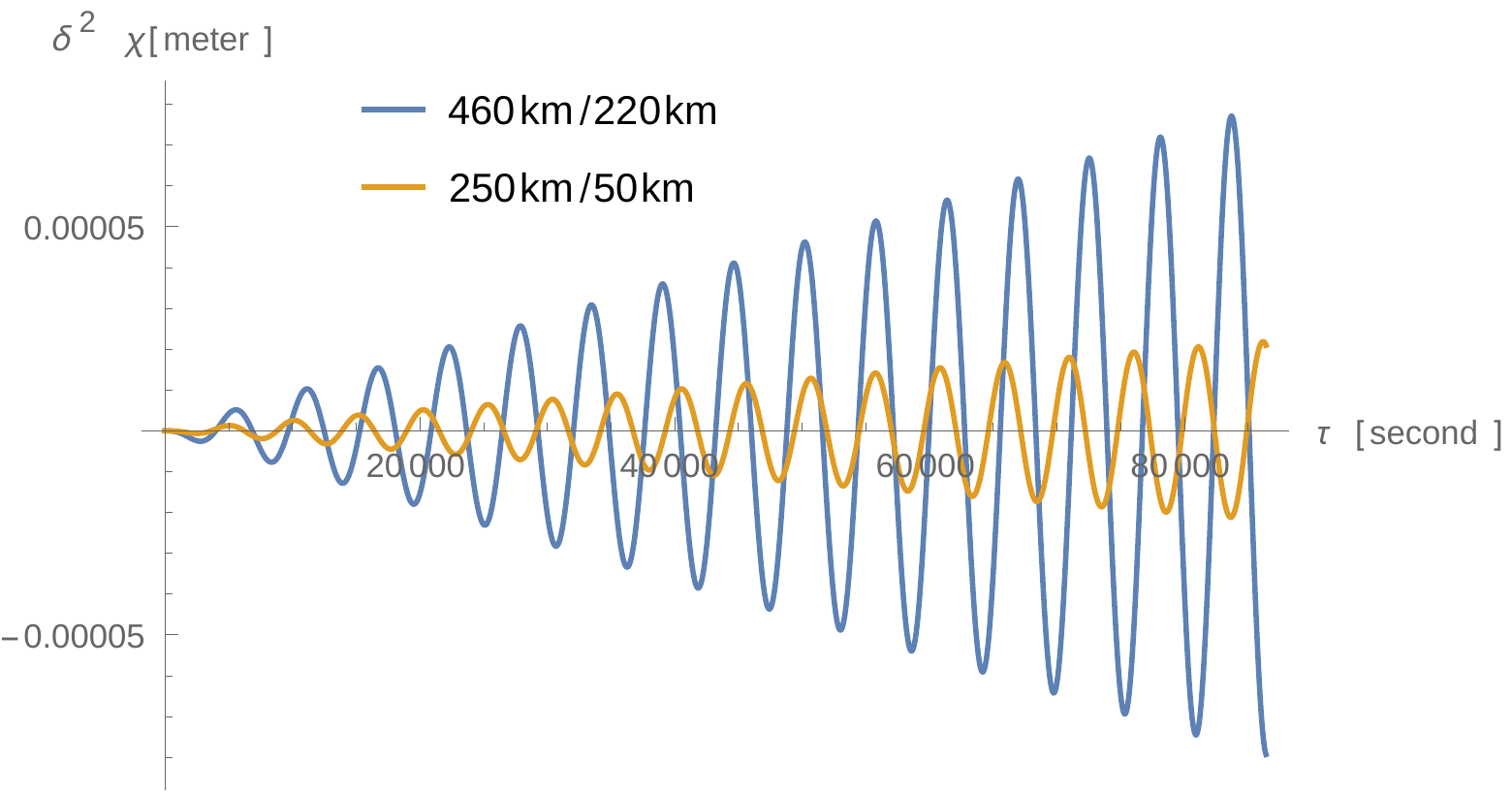}
\caption{The relative motions of the second satellite to the
  reference one in the radial direction. The orbits are chosen as near-circular
polar orbits with altitudes
  as $460km/250km$. The satellites separations are chosen as $220km/50km$.}
\label{fig:delta2}
\end{figure*}

 The solutions of
the equations of motions, that eq.(\ref{eq:inline}) and
eq.(\ref{eq:delta2}), with
general initial values
$\{\delta^1_0,\:\delta^2_0,\:\dot{\delta}^1_0,\:\dot{\delta}^2_0\}$
are
\begin{eqnarray}
\delta^1(\tau)&=& -\frac{\chi\rho_0 J  \sin i  \sin ( \omega
\tau )}{2 a^3} \tau-\frac{\chi\rho_0 J  \sin i( \cos (\omega\tau )-1)}{2 a^3\omega }\nonumber{}\\
&&+\delta^1_0
-\frac{2\dot{\delta}^2_0}{\omega}
- 3(\dot{\delta}^1_0+2\dot{\delta}^2_0\omega)
\tau+\frac{2\dot{\delta}^2_0\cos(\omega\tau)}
{\omega}\nonumber\\
&&+\frac{(4\dot{\delta}^1_0+6\delta^2_0)\sin(\omega\tau)}{
\omega } ,
\label{eq:fsd1} \\
\delta^2(\tau)&=&\frac{\chi\rho_0J\sin i\cos (\omega
\tau)}{4a^3}\tau -\frac{\chi\rho_0J\sin i\sin(\omega\tau)}{4a^3\omega}\nonumber{}\\
&&+\frac{2\dot{\delta}^1_0}{\omega}
+4\delta^2_0+\frac{\dot{\delta}^2_0\sin(\omega\tau)}{\omega}-\frac{(2\dot{\delta}
^1_0+3\delta^2_0
\omega)\cos (\omega\tau)}{\omega}.\label{eq:fsd2}
\end{eqnarray}
The most interesting signal in the range variations is the growing terms in
eq.(\ref{eq:fsd1}), which comes from the
in-phase actions of the CS perturbations in the along-track and the radial
directions. To be more specific, we write down the
solutions with the ideal initial conditions
$\delta^1_0=\delta^2_0=\dot{\delta}^1_0=\dot{\delta}^2_0=0$
and recover the SI units
\begin{eqnarray}
\delta \rho^{CS}(t)&=&\delta^1(t)=  -\frac{\chi G\rho_0 J  \sin i  \sin ( \omega t)}{2 c^2a^3} t-\frac{\chi G\rho_0 J  \sin i( \cos (\omega t )-1)}{2 c^2 a^3\omega }, \label{eq:sd1}\\
\delta^2(t)&=&\frac{\chi G\rho_0J\sin i\cos (\omega t)}{4c^2 a^3}t -\frac{\chi G\rho_0J\sin i\sin(\omega t)}{4c^2a^3\omega}\label{eq:sd2}.
\end{eqnarray}
Here we also replace the proper time $\tau$ of the reference satellite with the
coordinate time of the PN coordinates system, since the difference between these
two begins from $\mathcal{O}(\epsilon^2)t$ as showed in
eq.(\ref{eq:time}). In
figure.\ref{fig:planemotion}, figure.\ref{fig:delta1} and
figure.\ref{fig:delta2}, we
illustrate the above solutions
with the orbit options according to the GRACE and GRACE FO missions.
The length units in these figures are chosen as
$\chi \:meters$.

As discussed in the last subsection, the validities of the solutions
eq.(\ref{eq:fsd1}) and eq.(\ref{eq:fsd2}) are guaranteed under the
conditions that the deviations between the two satellites
should not exceed the 1PN level. Since the CS
parameter $\chi$ for Earth orbit satellites was already constrained to be a
rather small quantity $\chi\sim 0.17$ \cite{Smith2008}. Therefore, it will
take about $t\sim \frac{a}{\chi c \mathcal{O}(\epsilon^2)}\sim10^{7}yrs$ for
these deviations to reach $\delta^i\sim\rho_0\mathcal{O}(\epsilon)$,
$\dot{\delta}^i(\tau)
\sim \frac{\rho_0}{a}\mathcal{O}(\epsilon^2)$, and to break down the
above solutions. For experiments that carried by the SST missions with
at most 15 years life time, one needs not to
worry about this issue.

\section{The estimations of the measurement accuracy
and the concluding remarks\label{sec:measurements} }
From eq.(\ref{eq:sd1}), the magnitude of the CS range signal
$||\delta \rho^{CS}(t)||$ grows like
$\frac{\chi G \rho_0 J }{2 c^2 a^3}t$.
According to the orbits of the GRACE mission (see
table.\ref{tab:GRACE}) and the possible orbit choices of the
GRACE FO mission, we have $||\delta \rho^{CS}(t)||\sim 150N\chi\: \mu
m/day$ for the $460km$ altitude $220km$ separation option and
$||\delta \rho^{CS}(t)||\sim 40N\chi\: \mu
m/day$ for the $250km$ altitude $50km$ separation option, here $N$
denotes the days of the free flight. GRACE generally need $2\sim4$
orbits maneuvers per year, therefore the CS range signal accumulated in
each free flight will reach to a few $\chi cm$. As
mentioned in section.\ref{sec:GRACE}, the accuracy of the range measurements
near $10^{-4}Hz$ is about $1cm\sim2cm$ for GRACE \cite{Kim2002}. Therefore, with the
proper data analysis methods, such as matched filtering and etc., the data from
each free flight of GRACE (about $1.3\times10^7 seconds$) can in principle
set the constraint on the CS parameter as
$\chi\leq3.6\times10^{-4}$. From eq.(\ref{eq:mass}), and recovering the SI
units, we have
\[
M_{CS}=\frac{4\hbar c}{\chi a}.
\]
Therefore, the
length scale and the mass scale of the non-dynamical CS gravity to will be constrained as
\[
 32\pi\alpha\dot{\theta} \leq 2.5km, \:\:\ \ \ \:\:  M_{CS} \geq
3.1\times10^{-10}eV.
\]
The combination of the twelve years data of GRACE may improve the
constraint to $M_{CS}\geq 1.9\times10^{-9}eV$. For the
future GRACE FO mission that re-flies the GRACE orbits, the accuracy of the
range measurements around $10^{-4}Hz$ is about $100\mu m$ \cite{Sheard2012}. One
then expects the constraint from the data of one free flight to be
$\chi \leq 2.4\times10^{-6}$, which means that the length and mass scale of
non-dynamical CS gravity can be constrained to
\[
  32\pi\alpha\dot{\theta} \leq 0.017km, \:\:\ \ \ \:\:  M_{CS} \geq
4.6\times10^{-8}eV.
\]
For the GRACE FO mission that flies the $220km$ altitude $50km$ separation
option, the accuracy of the range measurements around $10^{-4}Hz$ is about $10\mu m$ \cite{Loomis2012}.  Then the constraints from one free flight will be
\[
32\pi\alpha \dot{\theta} \leq 6.0m, \:\:\ \ \ \:\:  M_{CS} \geq
1.3\times10^{-7}eV.
\]
The combination of the total nominal five years data of GRACE FO will
further improve the constraints for about $3\sim4$ times.

At last, we conclude this theoretical analysis with a brief discussions on the
corresponding data analysis procedure and the future plans following this
results. The SST missions such as GRACE and GRACE FO, designed originally for
satellite geodesies, may provide us the strongest tests and constraints on
the CS modified gravity up to now.
While, the corresponding data analysis procedure will form an non-trivial task,
since the frequency band around $10^{-4}Hz$ is affected by several noise
sources, such the Solar radiation pressure, Earth albedo, Earth atmosphere,
attitude disturbances and etc.. One needs to employ the data from
accelerometers, star sensors, magnetic torques and etc. to carefully remove the
non-conservative forces subjected to the spacecrafts. The range signals
produced by
Earth gravity multiples generally lie in the frequency band much higher than
the orbital frequency, and
can then be removed by proper low pass
filters. Take $J_2$ field for example, which is the
strongest multiple component producing signal with the lowest frequency. To
the leading order, the $J_2$ field will give rise
to a range signal along the nearly circular orbits
\[
 \delta\rho^{J_2}(t)=\frac{21G C_{20}\rho_0R^{2}\sin^{2}i\cos
(2\omega t)}{8a^ { 2}},
\]
which has twice the orbital frequency and can also be removed with proper
low pass filters. To summarize, one has to
start with the level 1b data in searching for the CS signals, which forms a
rather complicate task and will be left in future works.

\begin{acknowledgements}
This work was supported by the NSFC grands No. 11305255, No. 41104075 and
Central Universities funds (CHD2009JC100 and 2014G3262010).
\end{acknowledgements}

\appendix

\section{The standard PPN metric\label{sec:The-Standard-PPN}}

The standard PPN metric has the form \cite{Will2014}

\begin{eqnarray*}
g_{00} & = & -1+2U-2\beta U^{2}-2\xi\Phi_{W}+(2\gamma+2+\alpha_{3}+\zeta_{1}-2\xi)\Phi_{1}\\
 &  & +2(3\gamma-2\beta+1+\zeta_{2}+\xi)\Phi_{2}+2(1+\zeta_{3})\Phi_{3}+2(3\gamma+3\zeta_{4}-2\xi)\Phi_{4}\\
 &  & -(\zeta_{1}-2\xi)\mathcal{A}-(\alpha_{1}-\alpha_{2}-\alpha_{3})w^{2}U-\alpha_{2}w^{i}w^{j}U_{ij}+(2\alpha_{3}-\alpha_{1})w^{i}V_{i}+\mathcal{O}(\epsilon^{6}),\\
g_{0i} & = & -\frac{1}{2}(4\gamma+3+\alpha_{1}-\alpha_{2}+\zeta_{1}-2\xi)V_{i}-\frac{1}{2}(1+\alpha_{2}-\zeta_{1}+2\xi)W_{i}\\
 &  &
-\frac{1}{2}(\alpha_{1}-2\alpha_{2})w_{i}U-\alpha_{2}w^{j}U_{ij}+\mathcal{O}
(\epsilon^{5}),\\
g_{ij} & = & (1+2\gamma U)\delta_{ij}+\mathcal{O}(\epsilon^{4}),
\end{eqnarray*}
where the PN potentials read
\begin{eqnarray*}
U & = & \int\frac{\rho'}{|\mathbf{x}-\mathbf{x'}|}d^{3}x',\ \ \ \ \Phi_{1}=\int\frac{\rho'v'^{2}}{|\mathbf{x}-\mathbf{x'}|}d^{3}x',\\
\Phi_{2} & = & \int\frac{\rho'U'}{|\mathbf{x}-\mathbf{x'}|}d^{3}x',\ \ \ \ \Phi_{3}=\int\frac{\rho'\Pi'}{|\mathbf{x}-\mathbf{x'}|}d^{3}x',\\
\Phi_{4} & = & \int\frac{p'}{|\mathbf{x}-\mathbf{x'}|}d^{3}x',\ \ \ \ V_{i}=\int\frac{\rho'v'^{i}}{|\mathbf{x}-\mathbf{x'}|}d^{3}x',\\
W_{i} & = & \int\frac{\rho'[\mathbf{v}'\cdot(\mathbf{x}-\mathbf{x'})](x^{i}-x'^{i})}{|\mathbf{x}-\mathbf{x'}|^{3}}d^{3}x',\\
U_{ij} & = & \int\frac{\rho'(x^{i}-x'^{i})(x^{j}-x'^{j})}{|\mathbf{x}-\mathbf{x'}|^{3}}d^{3}x',\\
\mathcal{A} & = & \int\frac{\rho'[\mathbf{v}'\cdot(\mathbf{x}-\mathbf{x'})]^{2}}{|\mathbf{x}-\mathbf{x'}|^{3}}d^{3}x',\\
\Phi_{W} & = & \int\frac{\rho'\rho''(\mathbf{x}-\mathbf{x}')}{|\mathbf{x}-\mathbf{x'}|^{3}}\cdot(\frac{\mathbf{x'}-\mathbf{x}''}{|\mathbf{x'}-\mathbf{x}''|}-\frac{\mathbf{x}-\mathbf{x}''}{|\mathbf{x}-\mathbf{x}''|})d^{3}x'd^{3}x''.
\end{eqnarray*}
The matter variables are the rest mass density $\rho$, pressure $p$,
coordinate velocity of the matter field $v^{i}$, internal energy
per unit mass $\Pi$ and the coordinate velocity of the PPN coordinate
system relative to the mean rest-frame of the universe $w^{i}$. The
PN orders read
\[
v\sim\mathcal{O}(\epsilon),\ \ \ \ v^{2}\sim U\sim\Pi\sim\frac{p}{\rho}\sim\mathcal{O}(\epsilon^{2}).
\]

The standard PN parameters $\{\gamma,\ \beta,\ \xi\ ,\alpha_{1},\ \alpha_{2},\ \alpha_{3},\ \zeta_{1},\ \zeta_{2},\ \zeta_{3},\ \zeta_{4}\}$
have the following meanings. The parameters $\gamma$ and $\beta$
are the usual Eddington--Robertson--Schiff parameters used to describe
the ``classical'' tests of GR and are in some sense the most important
ones. For GR $\gamma=\beta=1$ are the only non-vanishing parameters.
The parameter $\xi$ measures the preferred-location effects, $\{\alpha_{1},\ \alpha_{2},\ \alpha_{3}\}$
measure the preferred-frame effects and$\{\alpha_{3},\ \zeta_{1},\ \zeta_{2},\ \zeta_{3},\ \zeta_{4}\}$
measure the violations of global conservation laws for total momentum.
The up-to-date values of these parameters are summarized in Tab.\ref{tab: PPN value}
\cite{Will2014}.
\begin{table*}
 \begin{center}
\begin{tabular}{ccc}
\hline
\textbf{Parameter}  & \textbf{Bound}  & \textbf{Experiment}\tabularnewline
\hline
\hline
$\gamma-1$  & $2.3\times10^{-5}$  & time delay in Cassini tracking\tabularnewline
\hline
 & $2\times10^{-4}$ & light deflection in VLBI\tabularnewline
\hline
$\beta-1$  &  $8\times10^{-5}$  & perihelion shift\tabularnewline
\hline
 & $2.3\times10^{-4}$ & Nordtvedt effect\tabularnewline
\hline
$\xi$  & $10^{-9}$  & spin precession of millisecond pulsars\tabularnewline
\hline
$\alpha_{1}$  & $4\times10^{-5}$ & orbital polarization of PSR J1738+0333\tabularnewline
\hline
 & $10^{-4}$  & Lunar laser ranging\tabularnewline
\hline
$\alpha_{2}$  & $2\times10^{-9}$  & spin precession of millisecond pulsars\tabularnewline
\hline
$\alpha_{3}$  & $4\times10^{-20}$  & pulsar spin down statistics\tabularnewline
\hline
$\zeta_{1}$  & 0.02  & combined PPN bounds\tabularnewline
\hline
$\zeta_{2}$  & $4\times10^{-5}$  & binary acceleration of PSR 1913+16\tabularnewline
\hline
$\zeta_{3}$  & $10^{-8}$  & Lunar acceleration\tabularnewline
\hline
$\zeta_{4}$  & --- & not independent\tabularnewline
\hline
\end{tabular}
\end{center}\protect\caption{Current values of PPN parameters.}
\label{tab: PPN value}
\end{table*}

\section{The Christoffel symbol and the tidal tensor
\label{sec:The-Christoffel-Symbol}}
The components of the Christoffel symbols
$\Gamma^{\mu}_{\:\:\:\:\rho\lambda}$ under the PN coordinate
system of section \ref{sec:CS} and the tidal matrix from the Riemann curvature
$K_{J}^{\:\:\ I}=R_{\lambda\nu\rho}^{\ \ \ \ \ \ \mu}T^{\lambda}T^{\rho}(\mathbf{e}_J)^{\nu}(\mathbf{e}^I)_{\mu}$
along the orbit eq.(\ref{eq:orbit}) are worked out as follows.

Here, we write down the complicate Christoffel symbols into matrix forms.
\begin{equation}
\Gamma^{0}_{\ \ \ 0\mu}=\frac{M (r-2 (\beta -1) M)}{r^4}
\left(
\begin{array}{c}
0\\
x^1\\
x^2\\
x^3
\end{array}
\right),\label{eq:G00m}
\end{equation}
\begin{equation}
\Gamma^{i}_{\ \ \ 0j}=
\left(
\begin{array}{ccc}
 0 & -\frac{\Delta J \left((x^1)^2+(x^2)^2-2 (x^3)^2\right) }{2 r^5} & -\frac{J
(6\Delta x^2 x^3 - \chi r x^1  )}{4 r^5} \\
\\
  \frac{\Delta J \left((x^1)^2+(x^2)^2-2 (x^3)^2\right) }{2 r^5} & 0 & \frac{J
(6\Delta x^1 x^3 +\chi r x^2  )}{4 r^5} \\
\\
\frac{J (6\Delta x^2 x^3 - \chi r x^1  )}{4 r^5}  & -\frac{J (6\Delta x^1 x^3 +\chi r x^2  )}{4 r^5} & 0 \\
\end{array}
\right)
,\label{eq:Gi0j}
\end{equation}
\begin{equation}
\Gamma^{1}_{\ \ \ ij}=-\frac{\gamma  M }{r^3}\left(
\begin{array}{ccc}
 x^1 & x^2 & x^3 \\
 x^2 & -x^1 & 0 \\
 x^3 & 0 & -x^1 \\
\end{array}
\right),\label{eq:G1ij}
\end{equation}
\begin{equation}
\Gamma^{2}_{\ \ \ ij}=-\frac{\gamma M}{r^3}
\left(
\begin{array}{ccc}
 -x^2 & x^1 & 0 \\
 x^1 & x^2 & x^3 \\
 0 & x^3 & -x^2 \\
\end{array}
\right),\label{eq:G2ij}
\end{equation}
\begin{equation}
\Gamma^{3}_{\ \ \ ij}=-\frac{\gamma M}{r^3}
\left(
\begin{array}{ccc}
 -x^3 & 0 & x^1 \\
 0 & -x^3 & x^2 \\
 x^1 & x^2 & x^3 \\
\end{array}
\right),\label{eq:G3ij}
\end{equation}
\begin{small}
\begin{eqnarray}
&&\Gamma^{0}_{\ \ \ ij}=\nonumber\\&&\left(
\begin{array}{ccc}
 \frac{3 \Delta J \left( x^2 (x^1)^3+ x^2
\left((x^2)^2+(x^3)^2\right)  x^1\right)}{ r^7} & \frac{3 \Delta J
\left(-(x^3)^2  (x^1)^2-(x^1)^4 +(x^2)^2
\left((x^2)^2+(x^3)^2\right)  \right)}{2 r^7} & \frac{3\Delta J \left(x^2
x^3  (x^1)^2+ x^2 x^3
\left((x^2)^2+(x^3)^2\right) \right)}{2 r^7} \\
+\frac{3 \chi J \left(3 x^3  (x^1)^2- x^3 \left((x^2)^2+(x^3)^2\right)
\right)}{2 r^6} & +\frac{6 \chi J
  x^2 x^3 x^1}{ r^6} & -\frac{\chi J \left(5
(x^1)^3+ \left(5 (x^2)^2-19 (x^3)^2\right)
x^1\right)}{4 r^6} \\
\\
 \frac{3 \Delta J
\left(-(x^3)^2  (x^1)^2-(x^1)^4 +(x^2)^2
\left((x^2)^2+(x^3)^2\right)  \right)}{2 r^7}& -\frac{3\Delta J
\left( x^2   (x^1)^3+ x^2
\left((x^2)^2+(x^3)^2\right)  x^1 \right)}{ r^7} & -\frac{3 \Delta J \left( x^3
 (x^1)^3+ x^3 \left((x^2)^2+(x^3)^2\right)
x^1 \right)}{2 r^7} \\
 +\frac{6 \chi J x^2 x^3 x^1}{ r^6} & -\frac{3\chi J
\left( x^3   (x^1)^2+ x^3 \left((x^3)^2-3 (x^2)^2\right)
\right)}{2 r^6} &  -\frac{\chi J \left(5 x^2 (x^1)^2+ x^2 \left(5 (x^2)^2-19
(x^3)^2\right)  \right)}{4 r^6}  \\
\\
 \frac{3\Delta J \left(x^2
x^3  (x^1)^2+ x^2 x^3
\left((x^2)^2+(x^3)^2\right) \right)}{2 r^7} & -\frac{3 \Delta J \left( x^3
 (x^1)^3+ x^3 \left((x^2)^2+(x^3)^2\right)
x^1 \right)}{2 r^7} & \frac{2 \chi J x^3
\left(-2 (x^1)^2-2 (x^2)^2+(x^3)^2\right)  }{r^6} \\
 -\frac{\chi J \left(5
(x^1)^3+ \left(5 (x^2)^2-19 (x^3)^2\right)
x^1\right)}{4 r^6}   &-\frac{\chi J \left(5 x^2 (x^1)^2+ x^2 \left(5 (x^2)^2-19
(x^3)^2\right)  \right)}{4 r^6} &   \\
\end{array}
\right).\nonumber\\
\label{eq:G0ij}
\end{eqnarray}
\end{small}

From Eq.(\ref{eq:gradient}), Eq.(\ref{eq:Z}), and Eq.(\ref{eq:e})-(\ref{eq:ie}),
the tidal matrix  $K^I_J=K^{N}+K^{GE}+K^{GM}+K^{CS}$
along the circular orbit in the Earth pointing local frame can be
worked out as
\begin{equation}
K^{N}=\frac{M}{a^{3}}\left(\begin{array}{cccc}
0&0&0&0\\
0&1 & 0 & 0\\
0&0 & -2 & 0\\
0&0 & 0 & 1
\end{array}\right),\label{eq:TN}
\end{equation}
\begin{equation}
K^{GE}=\frac{M}{a^{3}}\left(\begin{array}{cccc}
0&0&0&0\\
\\
0&-\frac{(2\beta+3\gamma-2)M}{a} & 0 & 0\\
\\
0&0 & \frac{(6\beta+5\gamma-5)M}{a}-(\gamma+2)a^{2}\omega^{2} & 0\\
\\
0&0 & 0 & \frac{(-2\beta-3\gamma+2)M}{a}+(2\gamma+1)\omega^{2}a^{2}
\end{array}\right),\label{eq:TGE}
\end{equation}
\begin{equation}
K^{GM}=\frac{J\omega}{a^{3}}\left(\begin{array}{cccc}
0&0&0&0\\
\\
0&0 & 0 & -\frac{3}{2}\Delta\sin i\cos\Psi\\
\\
0&0 & 3\Delta\cos i & \frac{9}{2}\Delta\sin i\sin\Psi\\
\\
0&-\frac{3}{2}\Delta\sin i\cos\Psi & \frac{9}{2}\Delta\sin i\sin\Psi & -3\Delta\cos i
\end{array}\right),\label{eq:TGM}
\end{equation}
\begin{equation}
K^{CS}=\frac{J\omega}{a^{3}}\left(\begin{array}{cccc}
0&0&0&0\\
\\
0&0 & -\frac{1}{4}\chi\sin i\sin\Psi & -\frac{1}{4}\chi\cos i\\
\\
0&-\frac{1}{4}\chi\sin i\sin\Psi & -\frac{3}{2}\chi\sin i\cos\Psi & 0\\
\\
0&-\frac{1}{4}\chi\cos i & 0 & \frac{1}{2}\chi\sin i\cos\Psi
\end{array}\right).\label{eq:TCS}
\end{equation}
Here, $K^{N}$, $K^{GE}$, $K^{GM}$ and $K^{CS}$
denote the gravitational tidal matrices from the Newtonian force,
the 1PN gravitoelectric force, the gravitomagnetic force and the contributions
from the CS modification.

\bibliographystyle{spphys}
\bibliography{Testing-Chern-Simons}

\end{document}